\documentstyle[psfig]{mn}

\newif\ifAMStwofonts
\ifoldfss
  \ifCUPmtlplainloaded \else
    \NewTextAlphabet{textbfit} {cmbxti10} {}
    \NewTextAlphabet{textbfss} {cmssbx10} {}
    \NewMathAlphabet{mathbfit} {cmbxti10} {} % for math mode
    \NewMathAlphabet{mathbfss} {cmssbx10} {} %  "   "    "
  \fi
  \ifAMStwofonts
    \ifCUPmtlplainloaded \else
      \NewSymbolFont{upmath} {eurm10}
      \NewSymbolFont{AMSa} {msam10}
      \NewMathSymbol{\upi}     {0}{upmath}{19}
      \NewMathSymbol{\umu}     {0}{upmath}{16}
      \NewMathSymbol{\upartial}{0}{upmath}{40}
      \NewMathSymbol{\leqslant}{3}{AMSa}{36}
      \NewMathSymbol{\geqslant}{3}{AMSa}{3E}

    \fi
  \fi
\fi % End of OFSS

\ifnfssone
  \newmathalphabet{\mathit}
  \addtoversion{normal}{\mathit}{cmr}{m}{it}
  \addtoversion{bold}{\mathit}{cmr}{bx}{it}
  \newmathalphabet{\mathbfit} % math mode version of \textbfit{..}
  \addtoversion{normal}{\mathbfit}{cmr}{bx}{it}
  \addtoversion{bold}{\mathbfit}{cmr}{bx}{it}
  \newmathalphabet{\mathbfss} % math mode version of \textbfss{..}
  \addtoversion{normal}{\mathbfss}{cmss}{bx}{n}
  \addtoversion{bold}{\mathbfss}{cmss}{bx}{n}
  \ifAMStwofonts
    \ifCUPmtlplainloaded \else
      \UseAMStwoboldmath
      \makeatletter
      \new@mathgroup\upmath@group
      \define@mathgroup\mv@normal\upmath@group{eur}{m}{n}
      \define@mathgroup\mv@bold\upmath@group{eur}{b}{n}
      \edef\UPM{\hexnumber\upmath@group}
      \new@mathgroup\amsa@group
      \define@mathgroup\mv@normal\amsa@group{msa}{m}{n}
      \define@mathgroup\mv@bold\amsa@group{msa}{m}{n}
      \edef\AMSa{\hexnumber\amsa@group}
      \makeatother
      \mathchardef\upi="0\UPM19
      \mathchardef\umu="0\UPM16
      \mathchardef\upartial="0\UPM40
      \mathchardef\leqslant="3\AMSa36
      \mathchardef\geqslant="3\AMSa3E
    \fi
  \fi
\fi % End of NFSS release 1

\ifnfsstwo
  \DeclareMathAlphabet{\mathbfit}{OT1}{cmr}{bx}{it}
  \SetMathAlphabet\mathbfit{bold}{OT1}{cmr}{bx}{it}
  \DeclareMathAlphabet{\mathbfss}{OT1}{cmss}{bx}{n}
  \SetMathAlphabet\mathbfss{bold}{OT1}{cmss}{bx}{n}
  \ifAMStwofonts
    \ifCUPmtlplainloaded \else
      \DeclareSymbolFont{UPM}{U}{eur}{m}{n}
      \SetSymbolFont{UPM}{bold}{U}{eur}{b}{n}
      \DeclareSymbolFont{AMSa}{U}{msa}{m}{n}
      \DeclareMathSymbol{\upi}{0}{UPM}{"19}
      \DeclareMathSymbol{\umu}{0}{UPM}{"16}
      \DeclareMathSymbol{\upartial}{0}{UPM}{"40}
      \DeclareMathSymbol{\leqslant}{3}{AMSa}{"36}
      \DeclareMathSymbol{\geqslant}{3}{AMSa}{"3E}
    \fi
  \fi
\fi % End of NFSS release 2

\ifCUPmtlplainloaded \else
  \ifAMStwofonts \else % If no AMS fonts
    \def\upi{\pi}
    \def\umu{\mu}
    \def\upartial{\partial}
  \fi
\fi

\title[Circumnuclear kinematics in NGC~5248]
        {Circumnuclear kinematics in NGC~5248: the origin of nuclear spiral arms}

\author[S. Laine et al.]
       {S. Laine$^{1,2}$\thanks{Present address: Space Telescope Science
       Institute, 3700 San Martin Drive, Baltimore, MD 21218, U.S.A.}, 
       J. H. Knapen$^{1,3}$, D. P\'erez--Ram\'\i rez$^1$\thanks{Present 
       address: Department of Physics, Michigan Technological University, 
       118 Fisher Hall, 1400 Townsend Drive, Houghton, MI 49931, U.S.A.}, 
       P. Englmaier$^4$ and M. Matthias$^5$\\
        $^1$Department of Physical Sciences, 
        University of Hertfordshire, College Lane, Hatfield, Herts AL10 9AB\\
	$^2$ Department of Physics and Astronomy, University of Kentucky,
	Lexington, KY 40506-0055, U.S.A.\\
	$^3$Isaac Newton Group of Telescopes, Apartado 321,
        E-35700 Santa Cruz de La Palma, Spain\\
        $^4$Max-Planck-Institut f\"{u}r extraterrestrische Physik, 85741 
	Garching, Germany\\	
	$^5$ Universit\"atssternwarte M\"unchen, Scheinerstr. 1, 81679 
	M\"unchen, Germany\\
	}
	
\vspace{-4cm}
	
\pagerange{\pageref{firstpage}--\pageref{lastpage}}
\pubyear{2000}

\begin{document}

\maketitle

\label{firstpage}

\begin{abstract}

We present for the first time a two-dimensional velocity field of the central
region of the grand-design spiral galaxy NGC~5248, at 0.9 arcsec spatial
resolution. The H$\alpha$ velocity field is dominated by circular rotation.
While no systematic streaming motions are seen in the area of the nuclear
grand-design spiral or the circumnuclear ring, the amplitude of residual
velocities, after subtracting a model circular velocity field, reaches 20
km~s$^{-1}$ in projection. The rotation curve levels out at around 140
km~s$^{-1}$, after a well-resolved and rather shallow rise. We have generated
an analytical model for the nuclear spiral and fitted it to our observations to
obtain estimates of the pattern speed of the spiral and the sound speed in the
central region of NGC~5248. Our results are consistent with a low pattern
speed, suggesting that the nuclear spiral rotates with the same rate as the
main spiral structure in NGC~5248, and thus that the spiral structure is
coupled from scales of a few hundred parsecs to several kiloparsecs. We have
also compared the observed structure and kinematics between the nuclear
regions  of NGC~5248 and M100. Several similarities and differences are
discussed, including the location of the peak emission regions on major and
minor axes, and the spiral arm streaming motions. We find no kinematic evidence
for a presence of a nuclear bar in NGC~5248.

\end{abstract}


\section{MOTIVATION}

During the past 40 years considerable progress has been made in
understanding the origin, nature and influence of spiral structure
in disc galaxies. The most commonly accepted theories now include
the density wave theory (Lin \& Shu 1964; Lin, Yuan \& Shu 1969) 
and its modal version
(Bertin et al. 1989a,b), which can generate spiral structures from
about 1 kpc to the outskirts of the disc. If dynamical (Lindblad)
resonances exist, the spiral structure in the collisionless
stellar component is permitted to exist between the inner and
outer resonances. The dissipative gas component is not subject to
these boundary conditions. Therefore, gaseous spiral structure may
extend to much smaller spatial scales.

The lack of spatial resolution both in observations and numerical models of
disc galaxies has hitherto hampered the study of spiral structure in the
central regions. The advent of near-infrared (NIR) and optical imaging  with
the {\it Hubble Space Telescope} ({\it HST}) and adaptive optics (AO)  on
4-meter class ground-based telescopes is now breaking the resolution  barrier,
allowing us to resolve galaxies within about 15 Mpc down to scales of 10 pc or
less. One of the main results from the increasing resolving power has been the
discovery of nuclear spirals (e.g., Ford et al. 1994; Rouan et al. 1998; Phillips
et al. 1996; Grillmair et al. 1997; Devereux, Ford \& Jacoby 1997;  Dopita et
al. 1997; Elmegreen et al. 1998;  Carollo, Stiavelli \& Mack 1998; Malkan,
Gorjian \& Tam 1998; Laine et al. 1999; Regan \& Mulchaey 1999; Martini \&
Pogge 1999). At the same time, the availability of powerful computers allows us
to create detailed numerical models which can resolve the spiral structure down
to 1--10 pc scales (e.g., Englmaier \& Shlosman 2000).

As an example of the recent breakthroughs, Laine et al. (1999) reported the
first detection of grand-design nuclear  spiral structure on scales of a few
hundred pc in the nuclear region of the spiral galaxy NGC 5248. In fact, spiral
structure exists in this galaxy on at least four different spatial scales,
extending from the few hundred pc scale minispiral to the 10 kpc scale weak,
outer arms. In between, there are the classical optically bright spiral arms,
and at the inner end of these, but outside the  circumnuclear starburst ring,
red spiral arms which are seen in NIR colour index maps. These arms
are reminiscent of the dust lanes often found along galactic bars. Since our
discovery of grand-design nuclear spiral arms, a few more have been
found, e.g., by Martini \& Pogge (1999; UGC 12138 and NGC 7682). A few galaxies
in the sample of Regan \& Mulchaey (1999) possess nuclear spiral dust lanes
reminiscent of grand-design spirals.

Spiral patterns in the nuclear region can be formed by at least
three processes.  First, a spiral density wave mode can grow in the inner disc.
In this case, the spiral density wave can exist between the Lindblad resonances
corresponding to the spiral pattern speed. Such a  wave would be formed by long
wavelength density waves in the stellar  and gaseous component. To avoid
perturbations from the dynamically  independent spiral pattern at larger
scales, the nuclear spiral pattern must exist inside the inner Lindblad
resonance (ILR) of the outer spiral  pattern, thus requiring a much higher
pattern speed for the nuclear spiral  pattern. Second, the spiral pattern can
be {\em driven} by the outer spiral pattern, as described and modelled by
Englmaier \& Shlosman (2000). In this case, the  potential well must be shallow
enough to prevent the spiral from becoming too tightly wound and destroyed by
dissipation. The pattern speeds  of the large and nuclear spirals are {\em
equal}. Finally, the nuclear spiral  may be formed by an acoustic instability
(Elmegreen et al. 1998), but this leads to irregular spiral patterns with
multiple arms.

We have obtained high resolution (0.9 arcsec), two-dimensional line of sight
velocity data for the central region of the spiral galaxy NGC~5248. In the
following sections we discuss the similarities and differences between the
H$\alpha$ morphology and kinematics in the core regions of NGC~5248 and M100,
and use the kinematical data, in  conjunction with the earlier high resolution
NIR colour index image (Laine et al. 1999), to obtain estimates of the
pattern speed of the nuclear spiral and the  sound speed of the ISM.

\section{OBSERVATIONS}

\begin{figure*}
\label{fig1}
\psfig{figure=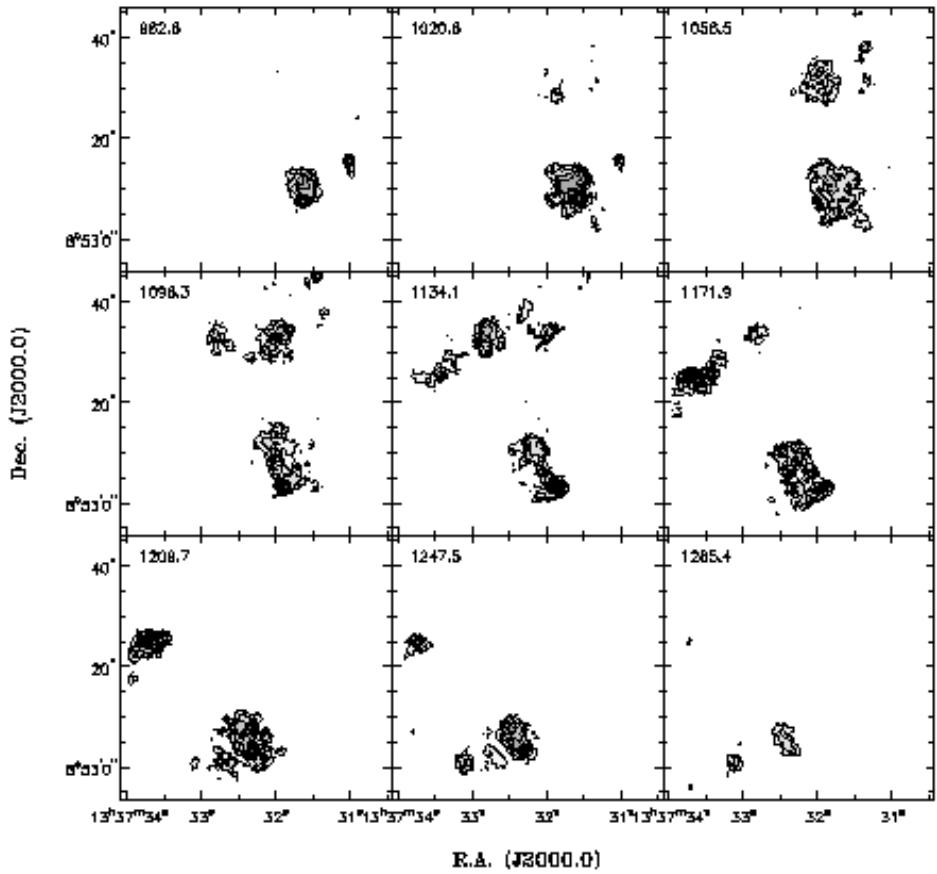,height=120mm}
\caption{Channel maps of the H$\alpha$ emission from NGC~5248 with about
1 arcsec resolution. Velocity of each channel is indicated in the upper left
corner. Only every second channel is shown. Contour and grey-scale levels are
approximately 3 to 32 $\sigma$, in steps of 3 $\sigma$.}
\end{figure*}

We used the TAURUS~II instrument in Fabry--P\'erot (FP) mode on the
4.2m William Herschel Telescope on La Palma during the nights of
1998 September 2 and 3. We windowed the TEK2 CCD camera to a size of
$600\times600$ pixels with a scale of 0.28 arcsec pixel$^{-1}$. The nights
were photometric with sub-arcsecond seeing. We used the
appropriately redshifted narrow-band H$\alpha$ filter ($\lambda_{\rm
c}$=6589\AA, $\Delta$$\lambda$=15\AA, using the galaxy's systemic velocity
$v_{\rm sys}$=1153~km~s$^{-1}$; NED) as an order-sorting filter. We
performed wavelength and phase calibration by observing a calibration
lamp before and after each science exposure. We subtracted the
background sky value from each separate plane, and shifted the planes to
the same position using fits to foreground stars. This produced two
data cubes of $300 \times$ 300 pixels $\times$ 55 `planes' in
wavelength, separated by 0.414 \AA, or 18.9 km\,s$^{-1}$.

\begin{figure*}
\label{fig2}
\psfig{figure=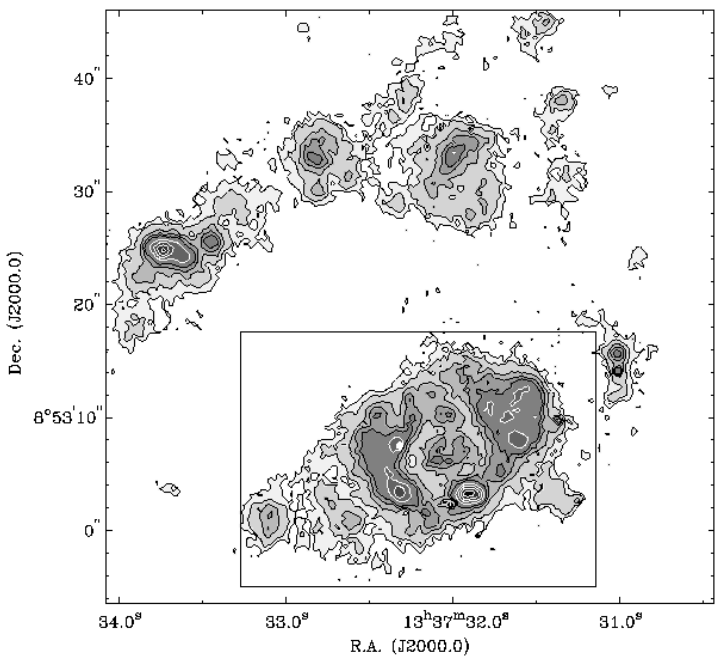,height=90mm,angle=270}
\caption{A grey-scale and contour image of the integrated H$\alpha$ emission 
in NGC~5248. The grey-scale and contour levels are 10, 25, 50, 75, 100, 150, 
200, 250 and 300 in arbitrary units. The rectangular box shows the area which 
is displayed in Fig.~3.}
\end{figure*}

\begin{figure*}
\label{fig3}
\psfig{figure=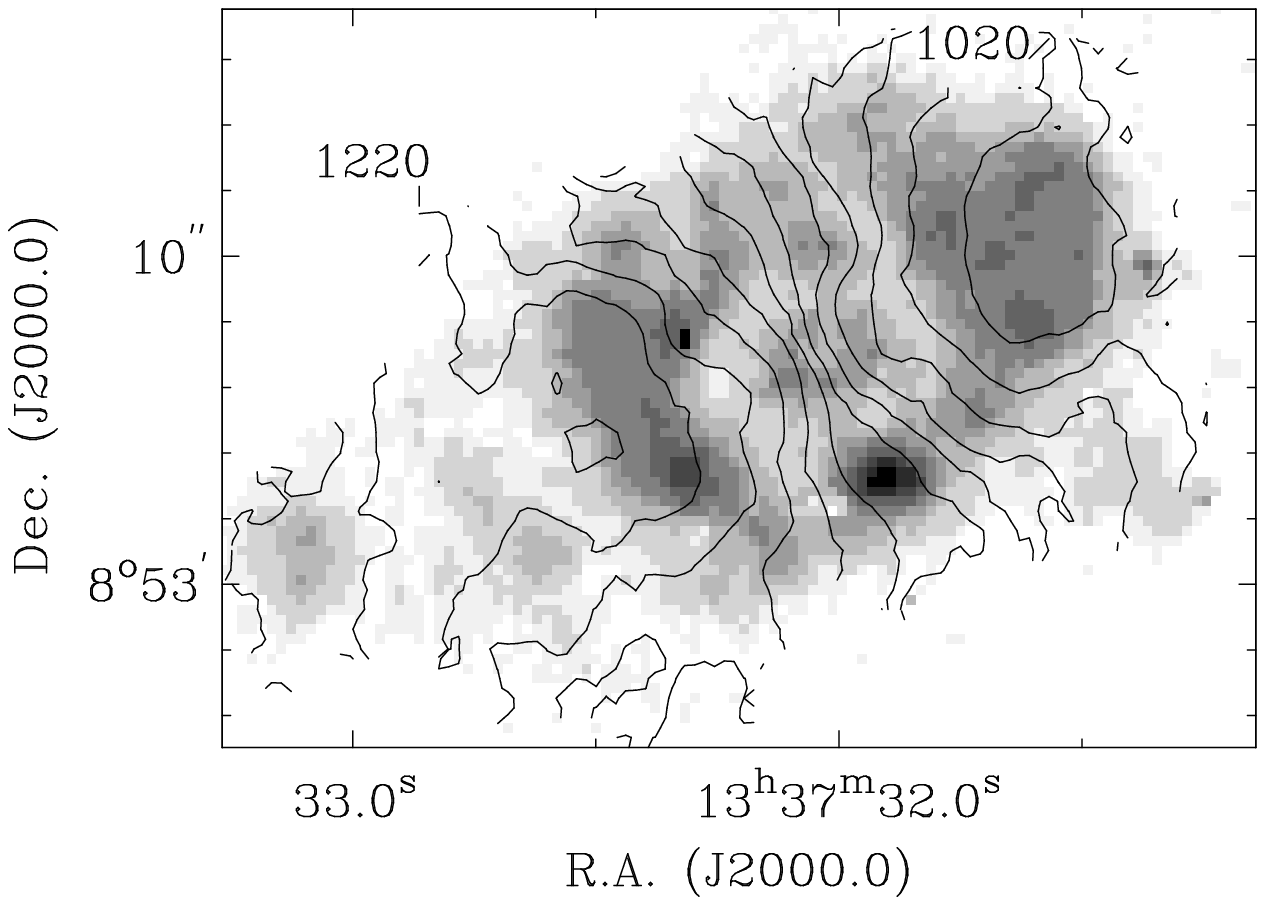,height=90mm,angle=270}
\caption{A grey-scale image of the integrated H$\alpha$ emission in NGC~5248
with line of sight velocity contours overlaid. The spacing of the velocity
contours is 20 km~s$^{-1}$ and two of the contours have been labelled.}
\end{figure*}

We combined the two individual datasets by placing the two cubes at the same
grid position using fits to the positions of foreground stars in the original
cubes and averaging the individual
planes in the data cubes. Astrometry was performed by comparing the
position of a bright foreground star to its position in the {\it
HST} Guide Star Catalog. 

\begin{figure*}
\label{fig4}
\psfig{figure=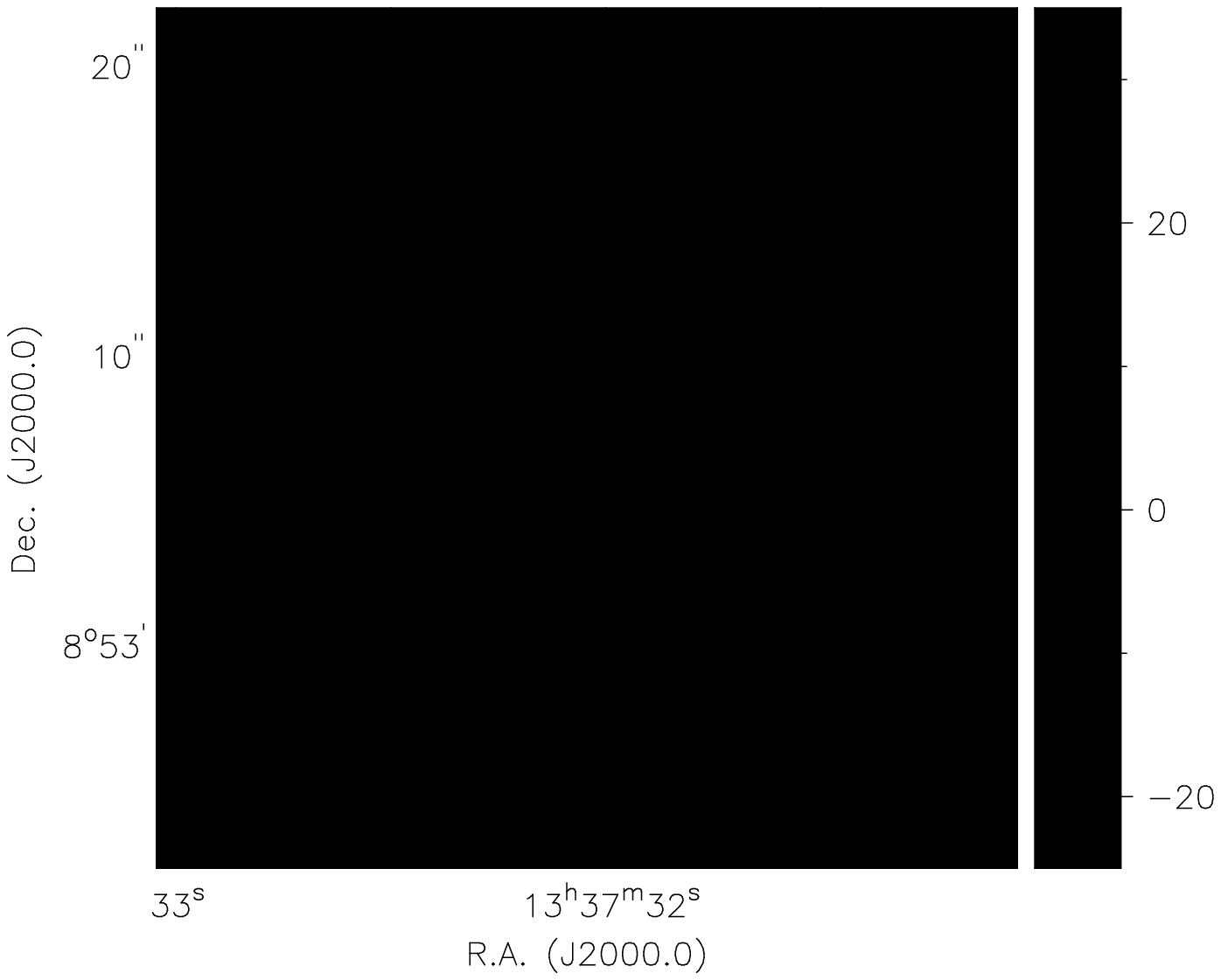,height=90mm}
\caption{A grey-scale representation of the residual velocity field after
subtracting the best-fitting model with purely circular rotation. The
wedge on the right-hand side indicates how differences in km~s$^{-1}$
correspond to grey-scales. Positive velocities indicate relatively higher
velocities in H$\alpha$.}
\end{figure*}

\begin{figure*} 
\label{fig5} 
\psfig{figure=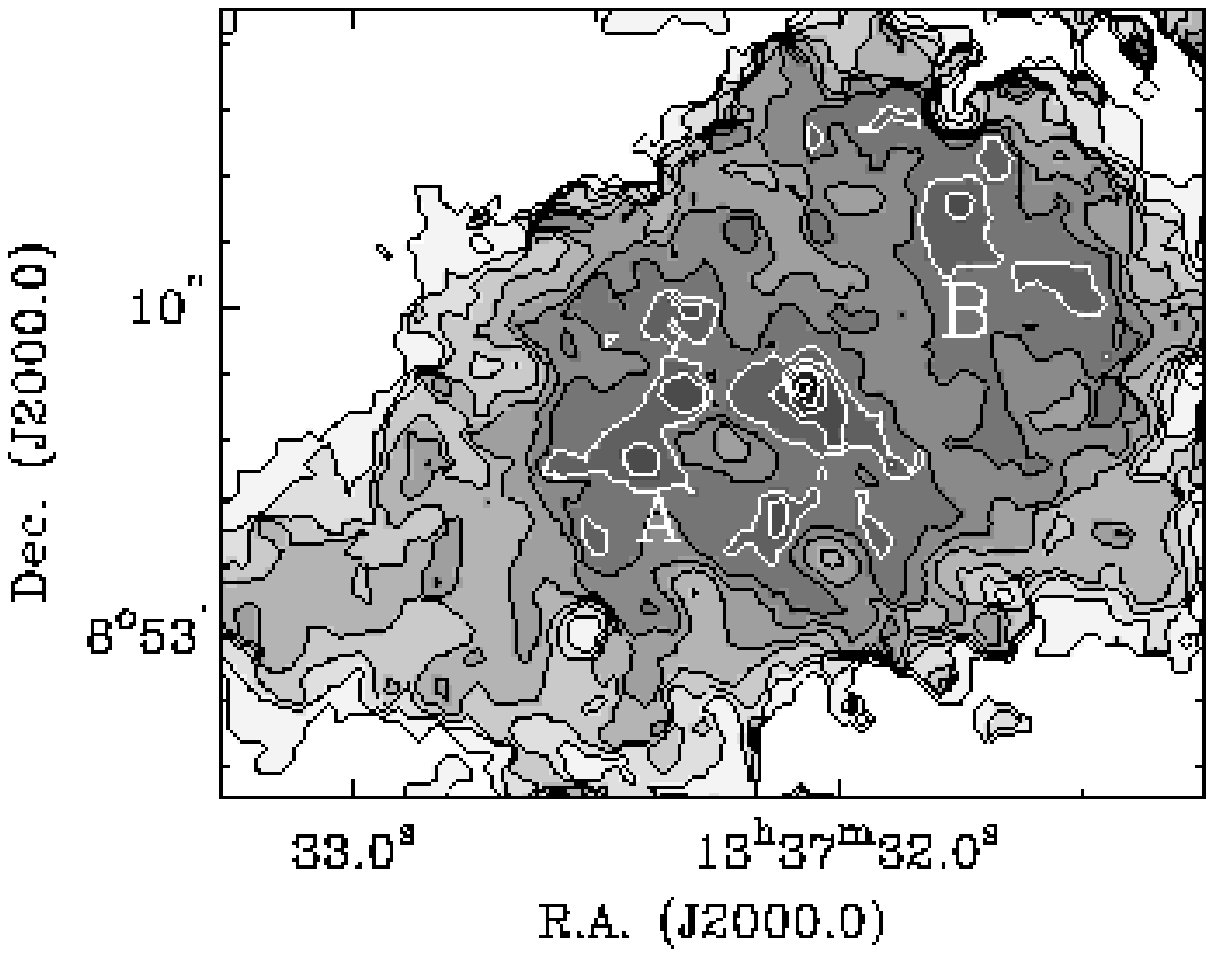,height=90mm,angle=270} 
\caption{A grey-scale and contour
representation of the smoothed second moment of the H$\alpha$ channel maps (2.5
arcsec resolution). The contours and grey-scales are shown at 10, 20, 25, 30,
35, 40, 45, 50, 55, 60, 65 and 70 km~s$^{-1}$. Two regions of enhanced velocity dispersion,
coinciding with the largest noncircular velocities in Fig.~3, are labelled with
`A' and `B'. The peak projected values in these regions reach 55~km~s$^{-1}$.} 
\end{figure*}

We determined which channels of the dataset were free of H$\alpha$ line
emission after smoothing the data cube to a resolution of 4~arcsec $\times$
4~arcsec. Subsequently, we subtracted the continuum emission after fitting the
continuum to the 24 line-free channels in the data cube. The H$\alpha$ emission
as a function of increasing wavelength, or velocity, is shown in a way
equivalent to the standard channel maps in radio astronomy in Fig.~1. Only the
circumnuclear area is shown, including the starburst ring.  

\begin{figure*}
\label{fig6}
\psfig{figure=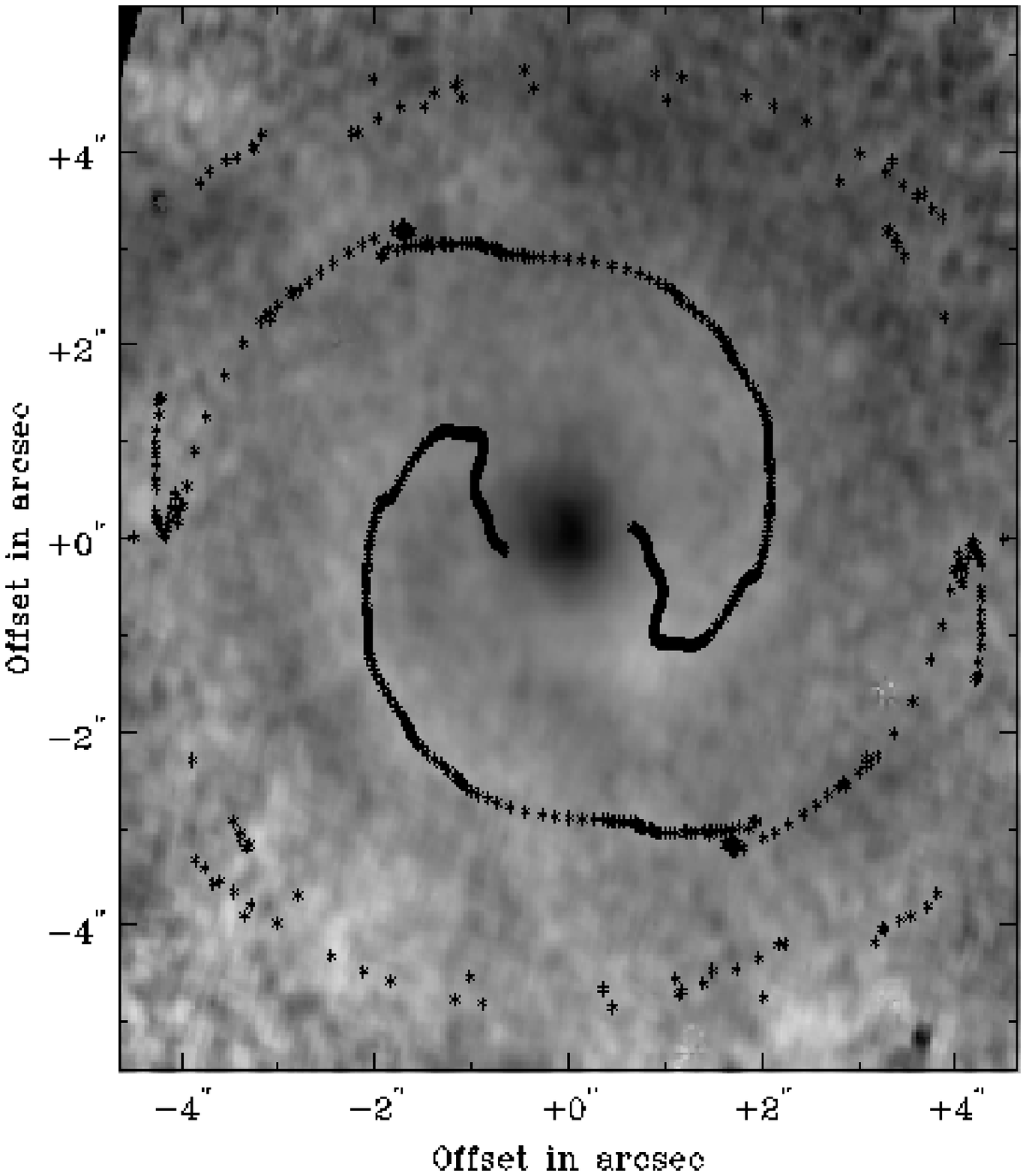,height=90mm,angle=270}
\caption{The J-K image from Laine et al. (1999). The image shows
the nuclear spiral and plotted on top, the phase angle of the m=2 multipole 
which we used to calculate the spiral pitch angle.}
\end{figure*}

We have performed both a Gaussian fitting of the spectra in the cube and a
moment analysis in the central region of NGC~5248. We choose to display the
moment maps here after comparing them to the map made by fitting Gaussians,
because the first moment map shows smoother velocity contours than the velocity
map derived by Gaussian fitting. We also made sure that the fitted spectra were
symmetrical  and single-valued.  The resulting moment 0 map (H$\alpha$ emission
distribution) is shown in Fig.~2. After smoothing the data cube to 2.5 arcsec
spatial resolution,  we created another set of moment maps. The resulting
smoothed velocity field is shown in Fig.~3. The detailed procedure used to
produce the moment maps is described in Knapen (1997) and Knapen et al. (2000).

\section{RESULTS}

\subsection{H$\alpha$ distribution}

\begin{figure*} 
\label{fig7} 
\psfig{figure=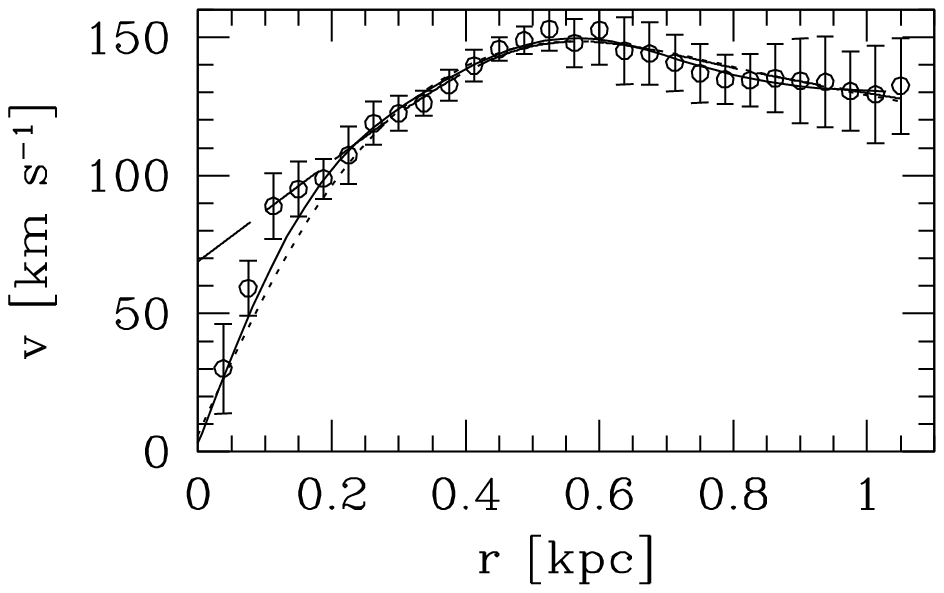,height=90mm} \caption{The
rotation curve (open circles) and Wahba \& Wendelberger method fits to the 
data points: excluding the two innermost data points (long-dashed), 
including all data points with low smoothing (solid) and  with higher 
smoothing (dotted).} 
\end{figure*}

The dominating feature in the H$\alpha$ distribution of Fig.~2 is the
circumnuclear ring with a major axis diameter of 14 arcsec. An {\it HST} image
of this ring (Maoz et al. 1996) shows that it consists of several 
tightly-packed `super star clusters'. Our H$\alpha$ image shows that inside the ring
at about 1 arcsec radius another pseudo-ring encircles the nucleus. This
pseudo-ring is probably associated with the dusty nuclear grand-design spiral
(Laine  et al. 1999). However, a detailed comparison with the dust spiral is
not possible because of the spatial resolutions that differ by almost a factor
of 10. We note, however, that the emission is clumpy, and that there is an
apparent break in the north side of both the main ring and the  nuclear
pseudo-ring. Since no reliable measures of the two-dimensional distribution of
extinction are available, we are unable to say whether such a break corresponds
to an asymmetric distribution of dust. Assuming that the spiral arms in
NGC~5248 are trailing, and considering that the southwest side of the ring is
receding with respect to the systemic velocity (see Section 3.2), the
north-eastern side  of the ring will be the far side if the ring is in the
plane of the main galaxy disc. Therefore, the optical depth may be larger in
the northern side, which would lead to weaker observed emission.

The most intense peaks of H$\alpha$ emission occur near the major
axis, towards southeast and northwest, and also on the minor axis
towards southwest. This situation is reminiscent of the morphology found in
M100 by Knapen et al. 2000. They found peaks near the major and minor axis
of the ring and argued that these locations represent regions of
crowding of the gas orbits in the vicinity of the inner inner Lindblad 
resonance (IILR).

Outside the circumnuclear ring, most of the detected emission comes
from the northern section of the main spiral arm, which is also bright in the
optical broad-band images of NGC~5248. Since we are mostly interested
in the circumnuclear region, we will not discuss the detected H$\alpha$
emission coming from the main optical spiral arms any further.

\subsection{H$\alpha$ kinematics}

\begin{figure*}
\label{fig8}
\psfig{figure=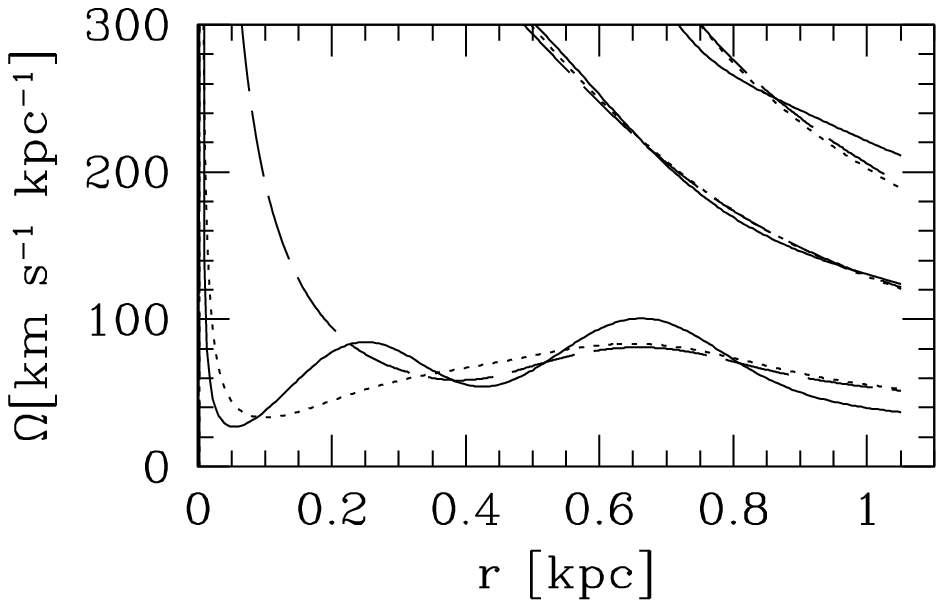,height=90mm}
\caption{Linear resonance diagram for the same fits as in Fig.~7. The
lower three curves correspond to $\Omega-\kappa/2$, the middle three curves to
$\Omega$, and the upper three curves show $\Omega+\kappa/2$, respectively.}
\end{figure*}

Figure~3 shows the velocity field of the nuclear region of NGC~5248, smoothed
to a resolution of 2.5 arcsec. The regular velocity contours in the Figure
indicate a relatively undisturbed velocity field without any strong noncircular
streaming motions. The velocity contours in the first moment image are closing
on both sides of the nucleus, implying that the rotation curve is declining. To
study the significance of noncircular (e.g., streaming) motions, we made an
image showing the difference between our best fitting purely circular rotation
model (see Section 4) and the observed velocity field. The resulting image is
shown in Figure 4. The largest regions of noncircular motion have amplitudes of
around 30 km~s$^{-1}$ (in the north-eastern corner) and 15 -- 20 km~s$^{-1}$
towards south and west. Deprojected, these velocities correspond to  deviations
from pure circular motions with a magnitude up to 50~km~s$^{-1}$. The regions
towards the northeast and west coincide with areas where spiral arms, coming in
from the outer disc, merge with the circumnuclear ring, and therefore they
could indicate the motions of gas coming in to the ring along the spiral arms.
Indeed, if we assume that the spiral arms are trailing, the northeastern side
of the galaxy is the far side, and negative residual velocities on that side
mean inflow, whereas the positive residual velocities towards the west also
imply inflow. This is consistent with the kinematical density wave theory which
predicts inward streaming motions along the spiral arms inside the corotation
resonance (Rohlfs 1977). This claim is supported by the smoothed (to 2.5
arcsec) second moment map (Fig.~5) which shows increased velocity dispersions
in two regions labelled `A' and  `B' in Figure~5. These two regions closely
coincide with the peak noncircular motions in Fig.~4 and with the peak
H$\alpha$ intensity. 

The lack of systematic noncircular streaming motions is strikingly different
from what was found in M100 (Knapen et al. 2000). There,
characteristic S-shaped contours of constant velocities were explained
with the help of the (nuclear) bar and additional velocity components
indicating streaming motions near the spiral armlets. An S-shaped signature is
not recognizable in Fig.~3, most likely because bars in NGC~5248 are absent
or weak (not recognized on images). However, there is a hint of
streaming motions connected with the incoming arms, as discussed above, towards
west-southwest at about 10 arcsec radius and perhaps towards east-northeast at
a comparable radius. The physical sense of these deviations was discussed
above. Therefore, the incoming spiral arms are most likely part of a
grand-design density wave system and not flocculent spiral arms.

The rotation curve was made from the unsmoothed data (with 0.9 arcsec
resolution) to avoid smoothing out the rise near the nucleus. We followed the
procedure of tilted ring fitting described by Begeman (1989). We used only
points within 60$^{\circ}$ of the kinematical major axis (to ignore radial
motions which are best seen near the kinematical minor axis) and cosine
$\theta$ weighting (where $\theta$ is the azimuthal distance from the
kinematical major axis in the plane of the galaxy). We fixed the inclination at
50$^{\circ}$, the position angle at 120$^{\circ}$, the systemic velocity at
1125~km~s$^{-1}$, and the position of the centre at  R.A. (J2000.0) = 13$^h$
37$^m$ 32\fs 05, Dec. (J2000.0) = 8\degr 53\arcmin 7\farcs 02, all obtained
from runs in which the parameter in question was left as a free parameter. The
rotation curve itself is shown in Figure 7, where the distances have been
converted into parsecs, assuming that 1 arcsec = 75 pc (obtained using a
distance of 15 Mpc, which in turn was obtained from the Hubble law with $H_{0}$
= 75 km~s$^{-1}$~Mpc$^{-1}$).

\begin{figure*} 
\label{fig9} 
\psfig{figure=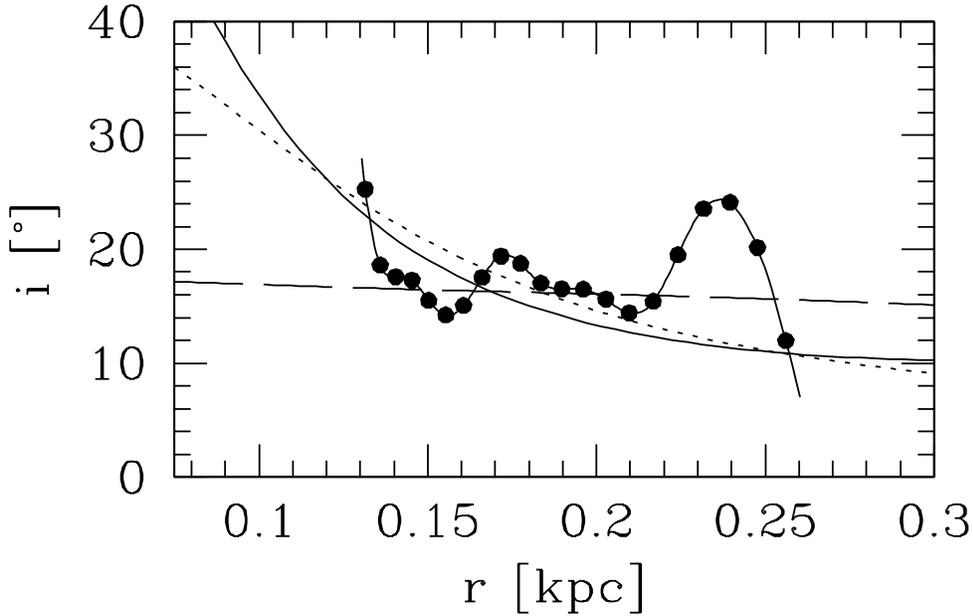,height=90mm}
\caption{Comparison of the measured spiral pitch angle seen in Fig.~6 (solid
line with filled dots) and the density wave theory prediction for a model
without the inner two points (dashed; $c_{\rm s}$=16~km~s$^{-1}$), a model with
all the points and a low smoothing (solid; $c_{\rm s}$=12~km~s$^{-1}$) and a
model with all the points and a higher smoothing (dotted; $c_{\rm
s}$=9~km~s$^{-1}$).} 
\end{figure*}

The rotation curve shows a relatively shallow rise, which is spatially  clearly
resolved. The peak rotation velocity occurs at about 145~km~s$^{-1}$ at 7
arcsec or 0.55 kpc distance from the centre and then the rotation velocity
starts a gradual decline. The shallow rise in NGC~5248 is in sharp contrast to
the rotation curve derived for M100 from similar Fabry--Perot observations by
Knapen et al. (2000), which most likely implies a smaller mass in the central
region of NGC~5248 compared to M100. There are hints that a shallow rise in the
rotation curve may be  directly related to the presence of nuclear spiral arms
(Knapen, Laine \& Rela\~no 1999). No recently observed large-scale rotation
curve is available for comparison with the kinematic behaviour of the
large-scale disk of NGC~5248. 

In order to obtain an estimate for the pattern speed of the spiral, we
generated an  analytical model using the description of Englmaier \& Shlosman
(2000). The following section gives a detailed description of the modeling.

\section{MODEL FOR THE NUCLEAR SPIRAL}

Recently, Englmaier \& Shlosman (2000) calculated analytical models, based on
the density wave theory, for nuclear spiral structure which is an extension of
the main spiral structure of a disc galaxy, inwards of the dynamical inner
Lindblad resonances (ILRs) which typically occur in the inner kiloparsec of a
galaxy. The main free parameters in the models are the pattern speed of the
spiral (or bar) and the sound speed of the interstellar medium (ISM), where the
sound speed is a  measure of the turbulent kinematic motions. Cowie (1980)
has  shown that this approach, which is wide-spread in the literature,  is in
fact a fair description of the ISM. Any given combination of those two
parameters, together with a specified rotation curve (or mass distribution)
will produce a model with a unique run of the spiral pitch angle. To test this
theory in the context of a grand-design nuclear spiral galaxy, we have used the
imaging data from Laine et al. (1999; Figure~6) and the kinematical data from
the current paper to investigate whether the pattern speed of the nuclear
spiral is consistent with the pattern speed of the main spiral structure at
larger radii, and whether the sound speed of the ISM is close to the values
measured at radii larger than 1 kpc in disc galaxies (typically $\sim$~10
km~s$^{-1}$).  Since the contrast of the observed dusty spiral arm pattern to
its background is extremely low (0.05 mag in the $J-K$ image of Laine et al.
1999), we assume that the mass of the nuclear spiral is very small and thus
the  self-gravity of the spiral arm can be neglected in our models.

To compare the density wave theory prediction with our kinematical (current
paper) and morphological (Laine et al. 1999) observations, we calculated the
pitch angle of the spiral arms as a function of radius in two ways. First, we
measured the pitch angle by eye-ball fitting a logarithmic spiral at the radii
where the pattern is well defined (from about 1.8 arcsec to 3.4 arcsec).
Second, we computed the phase angle for the $m=2$ mode as a function of radius.
While the former method concentrated on fitting the dark dust lanes, the latter
method tried to fit a sine curve to the azimuthal variation of the spiral in
the $J-K$ colour index image. The eye-ball fits were in agreement with the $m$=2
phase angle but less reliable. Therefore, in the following we only use the
latter fit. From the fitted phase angle we computed the pitch angle $i(r)$
shown in Fig.~9 by the solid line with filled dots. While we can recognize a
grand-design spiral between 0.13 and 0.25~kpc, a few perturbations  are visible
as waves in the $i(r)$ curve. At larger radii the spiral is lost in noise,
while at smaller radii the emission is dominated by the central bulge-like
component. As can been seen in Fig.~9, the pitch angle is in good agreement
with the theoretical prediction for all the considered fits to the rotation
curve data (the three model curves in Fig.~9). 

To calculate the density wave theory prediction we made a non-parametric fit to
the H$\alpha$ velocity data using the Wahba \& Wendelberger method (Wahba \&
Wendelberger 1980; Arnaboldi et al. 1998). With this algorithm we minimize the
quantity 

\begin{equation}
 \xi = {\rm N}^{-1} \sum_{{\rm i}=1}^{\rm N} ({\bf L}_{\rm i} {v} -
{{\tilde{v}}}_{\rm i}) \sigma_{\rm i}^{-2} + \lambda P({v}), 
\end{equation}

where N is the number of data points, ${\tilde{v}}_{\rm i}$ is the measured
rotation velocity of the i$^{\rm th}$ data point and $\sigma_{\rm i}$ is its
measurement error. The linear operator ${\bf L}$ decribes the Gaussian
seeing. The resulting intrinsic velocity field is represented by $v$;
${\bf L} {v}$ is the field as it would be observed with Gaussian
seeing. Therefore the first term of $\xi$ is a $\chi^2$ in the
observational space. The second term of $\xi$ is a regularising
functional in intrinsic space which prevents artificial amplification
of noise in the solution. The smoothing parameter $\lambda$ was
determined using a Fourier analysis of the recovered intrinsic velocity field.

The advantage of this method is that the result is not biased towards a
prescribed model. Individual errors in the rotation curve were taken into
account. Figure~7  shows three such model fits. The first fit (long-dashed
line) ignores the inner two points which are affected by the limited spatial
resolution. The other two fits were obtained using all the data points but also
taking into account a Gaussian-type point spread function corresponding to 0.9
arcsec seeing. Figure~7 shows the intrinsic rotation curves convolved with the
Gaussian seeing functional with a  low smoothing (solid line) and a higher
smoothing (dotted curve). Both are consistent with the Fourier-determined
smoothing parameter. From the intrinsic rotation curve, we calculated the
linear resonance diagram shown in Fig.~8. The use of a linear resonance diagram
in this galaxy is justified because a nonaxisymmetric bar perturbation in this
galaxy is relatively weak. Since the calculation of the epicyclic frequency
$\kappa$ requires the first derivative of $\Omega$, small bumps in the rotation
curve are amplified in the $\Omega-\kappa/2$ curve (or `ILR curve'). With
enough smoothing allowed in the Wahba \& Wendelberger method fit, we obtained a
smooth ILR curve without any bumps (see Fig.~8, solid and dotted curves). Since
the locations of ILRs are defined to be the radii where the ILR curves cross
the bar/spiral  pattern speed, the ILR curves in Fig.~8 are consistent with only one
ILR inside the one at 1.2 kpc found by Patsis, Grosb{\o}l \& Hiotelis (1997),
because the ILR curves rise inwards of 1.2 kpc and then dip only once to a
feasible bar pattern speed value ($<$ 20 km~s$^{-1}$). There may be another ILR
inside of 0.15 kpc but the spiral also disappears there. Therefore, the use of the
nuclear spiral theory developed by Englmaier \& Shlosman (2000) is justified.

Finally, we computed the wave number $k$ for the $m$~=~2 mode from the 
linearized dispersion relation of non-self-gravitating gaseous discs embedded
in an external frozen potential: 

\begin{equation}
(2\Omega(r)-2\Omega_{\rm P})^2=\kappa^2+k^2 v_{\rm s}^2 
\end{equation}
where $\Omega(r)=v/r$ is the circular velocity, $\Omega_{\rm P}$
the pattern speed, and $v_{\rm s}$ the sound speed. Note that near a Lindblad
resonance the dispersion relation is not applicable and formally the wave
number becomes zero. The wave number can then be used to derive the pitch angle
$i$ 
\begin{equation}
\tan i = |{2\over k r}|,
\end{equation}
 which will be independent of time.  

The resulting pitch angle depends on only two fit parameters: the pattern speed
and the sound speed $c_{\rm s}$ (Fig.~9). For the fit without the two innermost
points, the pitch angle stays mostly constant over the radii where the spiral
is observed. However, for fits which take all the points and the point spread
function into account, the spiral opens up closer to the centre, i.e., the
pitch angle increases rapidly. The best agreement with the observed pitch angle
was obtained when the sound speed was within a narrow range of
(9--16~km~s$^{-1}$), regardless of the inclusion or not of the innermost two
data points in the rotation curve or smoothing, and assuming that the pattern
speed (20~km~s$^{-1}$~kpc$^{-1}$, which after correcting for our distance of
15.4 Mpc instead of 10.5 Mpc as used by Patsis et al. becomes
13~km~s$^{-1}$~kpc$^{-1}$) found by Patsis et al. (1997) for the outer spiral
also holds for the nuclear spiral. Inside 1.5 arcsec (or 110 pc) the predicted
pitch angle depends strongly on the amount of smoothing applied and on how the
two innermost data points are treated.

\section{DISCUSSION AND CONCLUSIONS}

As recent high spatial resolution observations have shown, nuclear spiral
structure is abundant and at least in some cases takes on the morphology of
grand-design spiral arms (Laine et al. 1999; Martini \& Pogge 1999; Regan \&
Mulchaey 1999). Therefore, it is important to understand how such organized
structures can survive in the circumnuclear region where the dynamical time
scales are much shorter than in the outer disc. While flocculent spiral arms
may be explained with the help of the acoustic instabilities (Elmegreen et al.
1998), grand-design structure may only form either in an independent, rapidly
rotating component inside the inner ILR, or as a gaseous extension of the outer
spiral structure, thus forming part of the overall spiral pattern
that stretches from a few hundred pc scale up to several kpc. We can distinguish
between the latter two possibilities by measuring the pattern speed of the
nuclear spiral structure. NGC~5248 is a natural candidate for such a study.

In this paper we have shown that the circumnuclear velocity field of NGC~5248
is mostly regular, consistent with circular rotation, but inflow motion with
amplitude up to 50~km~s$^{-1}$ is seen near the points where the outer spiral arms
merge with the ring. Additional velocity perturbations are seen at the minor
axis, but only on one side of the galaxy. The peak H$\alpha$ emission areas
occur on the major axis and on the minor axis, similarly to what was observed
in M100. In contrast to M100, however, we do not see any systematic large-scale
streaming motions which could be associated with a nuclear bar or tightly
wound nuclear spirals. Bar streaming motions are probably absent because there
is no obvious bar in NGC~5248 at any observed spatial scale. The absence of
streaming motions in the ring may mean that the character of the ring is
different from that of M100, which possesses tightly-wound pseudo-spirals
that form a ring-like feature. Alternatively, it may reflect evolutionary
and possibly even age differences in the circumnuclear rings.  

We have also shown that the nuclear grand-design spiral structure may be driven
by one spiral pattern rotating at the same speed and extending from scales of
about 100 pc to several kiloparsecs. Galaxy structures which rotate with the
same pattern speed over a factor of ten in radial distance have been seen
before in barred spiral galaxies, where the bar and its associated ring system
rotate with the same pattern speed (e.g., ESO 507--16; Byrd, Ousley \& Dalla
Piazza 1998). Our model for the spiral pattern holds for a pattern speed of
13~km~s$^{-1}$~kpc$^{-1}$, which corresponds to the pattern speed derived by
Patsis, Grosb{\o}l \& Hiotelis (1997), after correction for the difference in
the galaxy distances that were used by them and by us. Specifically, high
pattern speed values do not agree with the observed arm pitch angle. The
acoustic spiral theory of Elmegreen et al. (1998) is not applicable to the
nuclear spiral structure of NGC~5248 since the spirals produced by that
mechanism are chaotic and consist of several branches ($m$ $>$ 2).

Our best fit for the sound speed of the ISM, 9--16~km~s$^{-1}$, rules
out any large increases in the sound speed in the nuclear few hundred
parsec region of NGC~5248 and is consistent with values derived for the
ISM in our Galaxy.

\section*{Acknowledgments}

This research has made use of the NASA/IPAC Extragalactic Database (NED) which 
is operated by the Jet Propulsion Laboratory, California Institute of 
Technology, under contract with the National Aeronautics and Space 
Administration. The William Herschel Telescope is operated on the island of La
Palma by the Isaac Newton Group in the Spanish Observatorio del Roque de los
Muchachos of the Instituto de Astrof\'\i sica de Canarias.

This paper has been typeseet from a \TeX/\LaTeX~file prepared by the
author.
\label{lastpage}

\end{document}